\documentstyle[psfig]{article}

\begin{document}

\title{On Accelerated Inertial Frames in Gravity and Electromagnetism}
\author{D. Lynden-Bell$^{1,2,3,4}$, J. Bi\v{c}\'{a}k$^{5,1,4}$ and
J. Katz$^{4,1}$} \maketitle

\centerline{Institute of Astronomy$^1$ and Clare College$^2$, Cambridge,} \centerline{ PPARC Senior Fellow on leave at School of Maths \& Physics,} \centerline {The Queen's University$^3$, Belfast.  BT7 1NN, }
\centerline{Racah Institute of Physics$^4$, The Hebrew University, Jerusalem,}
\centerline{Department of Theoretical Physics, Charles University$^5$, Prague.}

\begin{abstract}
When a charged insulating spherical shell is uniformly accelerated, an
oppositely directed electric field is produced inside.  Outside the
field is the Born field of a uniformly accelerated charge, modified by
a dipole.  Radiation is produced.

When the acceleration is annulled by the nearly uniform gravity field of an
external shell with a $1 + \beta \cos \theta$ surface distribution of
mass, the differently viewed Born field is static and joins a static
field outside the external shell; no radiation is produced.

We discuss gravitational analogues of these phenomena.  When a massive
spherical shell is accelerated, an untouched test mass inside
experiences a uniform gravity field and accelerates parallelly to
the surrounding shell.

In the strong gravity r\'egime we illustrate these effects using exact
conformastatic solutions of the Einstein-Maxwell equations with
charged dust.  We consider a massive charged shell on which the forces
due to nearly uniform electrical and gravitational fields balance.
Both fields are reduced inside by the ratio of the $g_{00}$ inside the
shell to that away from it.  The acceleration of a free test particle,
relative to a static observer, is reduced correspondingly.  We give
physical explanations of these effects.

\end{abstract}

\section{Introduction}

When a massive spherical shell is accelerated a free body inside it
experiences a parallel gravity field, i.e., the inertial axes inside
accelerate.  When the massive shell rotates the inertial axes inside
also rotate.  Weak versions of these inertia induction effects were
discovered by Einstein (1912) in Prague\cite{1} in the early
variable-velocity-of-light version of his gravitational theories.  He
described them clearly in his 1913 letter to Mach\cite{2}.  In
weak-field General Relativity he derives them in his book\cite{3}
but the non-Newtonian phenomena involved stretch the linearised theory
to a doubtful level of approximation.  Indeed the mass induction
effect also discussed by Einstein was demonstrated to be a coordinate
dependent effect by Brans (1962)\cite{4}.

Whereas a series of papers has been directed at the rotation of
inertial frames within spheres, etc., Thirring (1918,
1921)\cite{5}, Lense \& Thirring (1918)\cite{6}, Brill \& Cohen
(1966)\cite{7}, Lindblom \& Brill (1974)\cite{8}, Embacher
(1988)\cite{9}, Pfister \& Braun (1985)\cite{10}, Klein
(1993)\cite{11}, Lynden-Bell {\it et al}. (1995)\cite{12}, Katz {\it
et al}. (1998)\cite{13}, less attention has been given to linear
acceleration of the inertial frames within strongly relativistic
spherical shells.  This is perhaps because of the difficulty of
producing strong accelerations in General Relativity without severe
distortion of the problem to be solved\footnote{Recently the diploma
thesis of S.T. Hengge\cite{14} supervised by H. Pfister considered the
initial weak dragging within a highly charged massive spherical shell
placed between much smaller oppositely charged clouds.}.  Near
accelerating point masses the effects are complicated by the infall of
the inertial frames and the gravomagnetic fields due to motion, but a
few have been undeterred by these complications, e.g., Farhoosh \&
Zimmerman (1980)\cite{15}.

In continuation of our studies of Mach's Principle \& Machian effects
within General Relativity\cite{12,16,17,18,13}, we demonstrate here
the strong acceleration of the inertial frame {\it within} a massive
charged spherical shell of mass $m$ and radius $b$, which is
itself accelerated by a strong electric field.  To do this we bring
the system to rest by applying the equivalence principle and putting
the system in an external gravity field $g$.  Such a field can be
provided by a distant massive body $M$ which can also be so charged
that it supplies the electric field (Figure 1).  To ensure that the
principle of equivalence applies, the external tidal field of such a
body across our spherical shell must be negligible.  This can be
achieved at constant internal potential by shrinking both the shell
and its mass $m$ in proportion.  We are interested in effects of order
$g Gm/(bc^2)$ in acceleration.  We fix $g$ and $m/b$.  The tidal
accelerations are of order $(GM/R^3)b = g^{3/2} b (GM)^{-1/2}$ which
can be made as small as we like by decreasing $b$ (and $m$) at
constant ratio keeping $g$ and $M$ fixed.

The ratio of the accelerations we are interested in to those we wish
to neglect is then $(Gm/bc^2) (R/b)$ which we make as large as we like
by taking $b \ll R$ with $m/b$ fixed.  Notice we can do this with $g =
GM/R^2$ held fixed at any value we like so we may study {\em large}
accelerations and {\em strong} gravity $(Gm/bc^2 \sim 1)$.  We thus
reduce the problem to statics and use the beautiful conformastatic
solutions of Einstein's equations for charged dust with balancing
gravitational and electrostatic forces (see Appendix C).  We show that
the gravity field inside the massive charged spherical shell that sits
balanced in such applied fields is smaller than the applied gravity
field by the ratios of the $g_{00}$ which are themselves proportional
to $[1 + \sum Gm/(rc^2)]^{-2}$.  The problem may seem somewhat
contrived in that it is not met in nature.  We argue that it provides
an ideal test-bed for study of strong linear dragging by large
accelerations with strong relativistic effects and it is most unlikely
that a simpler thought experiment can be devised.  

However before entering the complications of general relativity it is
wise to study the analogous problems in the electrodynamics of
flat-space.  For them we can determine truly time-dependent phenomena
with the sphere initially at rest and only later given a uniform
acceleration $\alpha$.  For the transient behaviour see Appendix B.
We encounter the oft discussed question whether a uniformly
accelerated charge radiates and whether the equivalence principle
applies to a radiating charge.  This was well discussed by Fulton and
Rohrlich (1960)\cite{19} but even their excellent paper did not settle
the question for some (Rosen 1962\cite{20}).  We have found it
particularly interesting to draw the electric field lines of the Born
(1909) solution \cite{21} for a uniformly accelerated charge (see also
Schott\cite{22}) in several Lorentz frames including those in which
the charged particle is momentarily at rest; see Figures 2a, b, c, d,
e.  The field lines are circles that intersect at the particle (Bondi
\& Gold 1955\cite{23}).  We join them far away to the initial Coulomb
field of the fast moving particle before the steady deceleration
brought it to near rest.  In that field there is also a magnetic field
with circular magnetic field lines around the direction the particle
was moving.  Such figures help us to envision how the Born solution
describes the radiating field of an accelerated particle in the one
case but can also describe the non-radiative static field of a
particle held stationary in a uniform gravity field!  As Boulware
(1980)\cite{24} showed, the principle of equivalence still holds in
the region accessible to both the static and the coaccelerated
observer, but does not hold beyond the event horizon of the
coaccelerated observer who can see neither the radiation of the charge
that is accelerated with him nor its effects.

\section{The field of an accelerated charged spherical shell}

For comparison to our sympathetic dragging calculations in gravity, we
need the electromagnetic field inside a uniformly charged accelerated
shell.  Whereas the Born field of a {\em point} charge is well
discussed, we give here the corresponding field inside and outside an
accelerated charged {\em shell}.  Inside the field rapidly $(2b/c)$
settles down to a uniform one but outside the propagation takes
longer.  Initially we calculate the field for short times only but the
near field outside soon becomes that of a charge with a forward
pointing electric dipole and no higher multipoles.  These fit
respectively the exact Born field and the exact field of uniformly
accelerated dipole, both of which are known\cite{25}.  Thus our
short-time calculation of the dipole moment tells us the coefficient
of the dipole solution which should be added to the Born field to give
the complete solution for the uniformly accelerated shell at all times.

We shall consider times close to the moment when the shell is
stationary in our frame.  Both charges and currents lie close to the
shell $r=b$.  The surface charge density is $Q/(4 \pi b^2)$ so
we have a charge density
$$\rho = {Q \over 4 \pi b^2} \left( 1 - D {\partial \over \partial z}
\right) \delta (r - b) \ , \eqno (2.1)$$
where $D(t)$ is the small displacement of the sphere at time $t$.
Likewise the current on the sphere is
$${\bf j} = {Q {\hat {\bf z}} \over 4 \pi b^2c} \dot D \left( 1 - D {\partial
\over \partial z} \right) \delta (r - b) \ . \eqno (2.2)$$
Hats denote unit vectors.  

To find the potentials we wish to solve
$$\Box^2 \Phi = - 4 \pi \rho \ , \eqno (2.3)$$
$$\Box^2 {\bf A} = - 4 \pi {\bf j} \ . \eqno (2.4)$$
We first solve for the Green's function $\chi$ such that
$$\Box^2 \chi = - \delta (t - t') \delta (r -b) \ . \eqno (2.5)$$
Then $$\Phi = {Q \over b^2} \int \left( 1 - D (t') {\partial \over
\partial z} \right) \chi dt' \ , \eqno (2.6)$$
and $${\bf A} = {Q {\hat {\bf z}} \over b^2 c} \int {\dot D} (t') \left( 1 - D
(t') {\partial \over \partial z} \right) \chi d t' \ . \eqno (2.7)$$
Now $${1 \over r} \left( {\partial^2 \over \partial r^2} - {1 \over
c^2} {\partial^2 \over \partial t^2} \right) (r \chi) = - \delta (t -
t') \delta (r - b) \ . \eqno (2.8)$$

The Green's function $\chi$ which solves (2.5) and has only a retarded
solution in $r \geq b$ is readily shown to be

$$\chi = - {bc \over 2r} \left\{ {\Theta \left( t - t' - {r +b
\over c} \right) - \Theta \left( t - t' + {r - b \over c} \right)
\ , \ r \leq b \atop \Theta \left( t - t' - {r+b \over c} \right)
- \Theta \left( t - t' - {r - b \over c} \right) \ , \ r \geq b}
\right. \eqno (2.9) $$
where ${\Theta}$ is 1 for a positive argument and zero
otherwise.
A detailed derivation of (2.9) is given in Katz {\it et al}. (1998)\cite{13}.

Returning to (2.6) and using ${\partial \over \partial z} f (r) = {z
\over r} {\partial \over \partial r} f$ we find
$$\Phi = Q b^{-2} \left[ \int ^\infty _{- \infty} \chi dt' - {z \over
r} {\partial \over \partial r} \int ^\infty _{- \infty} D (t') \chi
(t') dt' \right] \ . \eqno (2.10)$$
For a uniformly accelerated sphere, $D (t') = {\textstyle{1 \over 2}}
\alpha t'^2$, and the integrals are readily evaluated to give
$$\Phi = \left\{ {Qb^{-1} \left( 1 - {1 \over 3} \alpha c^{-2} z
\right) \quad \ \, {\rm inside, \ i.e.,} \ |{\bf r} - {1 \over 2}
\alpha t^2 {\hat {\bf z}}| \leq b \atop Q r^{-1} \left \{ 1 + {1 \over
2} \alpha c^{-2} z \left[ \left( c^2 t^2 + {1 \over 3} b^2 \right)
r^{-2} - 1 \right] \right \}, \ \ r \geq b } \right. \ . \eqno (2.11)$$

Returning to (2.7) we see that the $\dot D D$ term is quadratic in the
acceleration $\alpha$ so the first term will suffice when $\alpha b
c^{-2} \ll 1$.  For an acceleration of 1g this means $b$ much smaller
than a light year $\sim 10^{18}$ cm!  Evaluating ${\bf A}$ for a
general $D$ we find
$${\bf A} = - {\textstyle{1 \over 2}} Q {\hat {\bf z}} b ^{-1} r^{-1}
\left \{ {D \left( t - {r + b \over c} \right ) - D \left( t + {r - b
\over c} \right) \qquad r \leq b\, , \atop D \left( t - {r + b \over c}
\right) - D \left( t - {r - b \over c} \right) \qquad r \geq b\, , } \right
. \eqno (2.12)$$
or for $D = {\textstyle{1 \over 2}} \alpha t^2$
$${\bf A} = Q \alpha c^{-2} {\hat {\bf z}} \left \{ {b^{-1} (ct - b)
\qquad r \leq b \atop r^{-1} (ct - r) \qquad r \geq b} \right. \
. \eqno (2.13)$$
The fields are readily derived from ${\bf E} = -c^{-1} {\dot {\bf A}} -
\mbox {\boldmath $\nabla$}  \Phi$.  Inside the uniformly accelerated
sphere we have, writing $a = c^2/\alpha$, 
$${\bf E}_{\rm in} = - {2 \over 3} Q (a b)^{-1} {\hat {\bf z}} \qquad r \leq b \
, \eqno (2.14)$$
and outside
$$\displaylines {{\bf E}_{\rm out} = Q r^{-2} \left\{ 1 - {\textstyle {1 \over 2}} (z/a)
\left [ 1 - r^{-2} (3 c^2 t^2 + b^2) \right] \right \} {\hat {\bf r}}
- \cr \hfill - {\textstyle{1 \over 2}} Q (ar)^{-1} \left [ 1 + {\textstyle {1 \over 3}} r^{-2} (3c^2
t^2 + b^2) \right] {\hat {\bf z}} \ \ \ \  , \ \ \ \ \sqrt{ab} \gg r \geq b \
. \hfill {(2.15)} \cr}$$

We have performed our calculation under the assumption that the
movement of the sphere during the time considered is much less than
the radius of the sphere.  Thus the result is only valid provided the
times involved obey
$${\textstyle {1 \over 2}} \alpha t^2 \ll b \ . $$
During this time the signal will propagate to $r = ct$ so the solution
is only valid where $r^2 \ll 2 b c ^2 \alpha ^{-1} = 2 ab$.  To get
the field at greater distances we may refer to Born's solution for the
field of a point charge which is uniformly accelerated in its own
frame so its position is given by 
$$z = \sqrt{a^2 + c^2 t^2} - a = {\overline z} - a \ . \eqno (2.16)$$
This is given in Appendix A.

We initially expected the {\it external} field of our accelerated
sphere on which the charge density was frozen, to match exactly the
``small acceleration'' (large $a$) limit of the Born solution.
However the presence of $b^2$ terms in (2.15) shows that this cannot
be the case since in Born's solution there is no $b$.  Indeed those
terms show that our accelerated sphere has a dipole of moment ${\bf P} =
{\textstyle {1 \over 6}} Q (b^2/a) {\hat {\bf z}}$.  Bi\v{c}\'{a}k and
Muschall (1991)\cite{25} showed how to get the exact solutions for
uniformly accelerated multipoles by differentiating Born's solution
with respect to $a$.  Putting their dipole equal to the above ${\bf
P}$ and adding the result to Born's solution gives the exact external
solution for our uniformly accelerated sphere, see (2.17),
etc. below.  Of course if we set $b =t =0$ our external fields fit the
Born solution including terms of order $1/a$ but not when $r>a$.  The physical origin of our forward pointing dipole is
interesting.  Our ${\bf E}$ field arises as it should from a
spherically symmetrical distribution of surface charge.  In the Born
solution the field-lines droop under their acceleration-induced
weight (see Fig. 2b).  By demanding that our sphere be of uniform
surface density we have demanded no droop out to $r = b$.  The droop
of the Born field lines is therefore offset by the forward pointing
dipolar field we have found.

If we cut Born's solution at $r = b$ we may replace the inside with a
charged shell but, its internal ${\bf E}$ field is not our $-
{\textstyle{2 \over 3}} Q (a b)^{-1} {\hat {\bf z}}$ but another
uniform field whose coefficient is $-{\textstyle {1 \over 2}}$ in
place of $- {\textstyle {2 \over 3}}$.  The reason for this difference
is that the surface charge density on a sphere that gives the Born
field outside is {\it not} uniform but rather $\sigma = {Q \over 4 \pi
b^2} (1 - {\textstyle {1 \over 2}} {b \over a} \cos \theta)$.  It is the internal
field of this non-uniform charge density with excess charge at the
bottom that reduces the internal field strength from ${\textstyle{2
\over 3}}$ to ${\textstyle{1 \over 2}}$.  Thus our `exact' solution for
a uniformly charged uniformly accelerated sphere is not Born's field
outside, but is given by 
$${\left. \Phi = \left[ 1 + {1 \over 3} b^2 \left({\partial \over \partial
a^2} \right)_{{\overline z}} \right ] \Phi^B (R, {\overline z}, t, a^z )
\atop \ \ {\bf A} = {\hat {\bf z}} \left[ 1 + {1 \over 3} b^2 \left(
{\partial \over \partial a^2} \right)_{{\overline z}} \right] A^B (R,
{\overline z}, t, a^2 ) \right\}} \ , \eqno (2.17)$$
where, to allow steady uniform acceleration, we have to take $b \ll a
= c^2/ \alpha$, and $\Phi ^B  $ and $ A^B$ are given in Appendix A.

Inside the sphere, which becomes Lorentz contracted into an oblate
spheroid away from $t = 0$, the electric field is always given by
(2.14) with no magnetic field there.  These properties follow from
Lorentz transformation of our $t = 0 $ field.  We discuss the
transients when the shell starts from rest in Appendix B.  

Born's field is static in uniformly accelerated axes.

\section{Gravity in Accelerated Shells}

In weak field theory, Einstein considered a spherical shell of mass
$m$ and radius $b$ that was weakly accelerated by some unspecified
forces and looked at the metric inside the shell.  We wish to
accelerate a massive shell quite rapidly and we need to specify some
way of doing it because the stresses involved will produce their own
gravity.  We therefore consider a uniformly charged insulating massive
spherical shell in an electric field so that every element will
accelerate uniformly without any acceleration-induced stresses.
Einstein using his 1912 theory\cite{1} found a uniform gravity field
of strength ${fGm \over c^2b} \, \mbox{\boldmath $\alpha$}$ within his
accelerating shell, where $f$ was $3/2$ using coordinate time.
However in his 1955 book\cite{3} he gives the weak field relativistic
formulae in the de Donder gauge.  Evaluating these, including
retardation in the potential, gives an ${\dot {\bf A}}$ term -- $4
\times$ the analogous electrical one and a ${\mbox{\boldmath
$\nabla$}} \Phi$ term equal and opposite to the electrical one.  So we
get from Einstein's weak field formulae $f=11/3$.  So we should expect
a free test body inside our shell to accelerate in sympathy but less
rapidly than the shell itself.  However time-dependent problems in
general relativity are difficult, so rather than solving a difficult
problem we shall use the principle of equivalence to reduce it to an
easy one.  The acceleration of the whole shell is equivalent to a
uniform gravity field imposed from outside.  If ${\bf E}$ is the
accelerating electric field and $q$ and $m$ are the charge and mass of
the shell, then the upward acceleration is $\mbox{\boldmath $\alpha$}
= q {\bf E}/m$ as shown in Bi\v{c}\'{a}k 1980\cite{26}.  The extra
`fictitious' downward gravity field that appears to a coaccelerated
observer is ${\bf g} = - \mbox{\boldmath $\alpha$} $, and the
principle of equivalence is that the metric felt by such an observer
will be the same as that static metric felt when the extra gravity
field, ${\bf g}$, is really present.  To generate such a real gravity
field we place a very large mass $M$ at a large distance $Z$ down the
z axis such that the gravity field due to it is just ${\bf g}$ that is
$g \simeq GM/Z^2$.  We may also use this mass to generate the electric
field ${\bf E}$ by giving it a charge $Q = E Z^2$.  Then the
gravitational attraction of $m$ by $M$ exactly cancels their
electrical repulsion so the whole system is static and at least
classically $qQ = GmM$.  Because it makes even the general relativity
easy we shall take both the sphere and the large $M$ to have equal
charge to mass ratio so $q/m = Q/M = \sqrt{G}$.  That is each has the
charge/mass ratio of the extreme Reisner-Nordstr\"om black hole.  To
attain bodies with $Q = \sqrt{G}M$ we put together $N(1-f)$ hydrogen
atoms and $Nf$ protons to give $Q = fNe$ and $M \simeq Nm_H$.  Then
$QG^{-1/2} M^{-1} = e G^{-1/2} m^{-1}_H f = f/9 \times 10^{-19}$, so we must
choose $f = 9 \times 10^{-19}$ to get balanced electrical and gravitational
forces.  This gives about $10^{18}$ neutral hydrogens per proton,
implying a charge of 1 Coulomb on $1.18 \times 10^{13}$gm or about
$1.7 \times 10^{20}$ Coulombs on a solar mass which would give it a potential
of $2.5 \times 10^{19}$ volts, far more than realistic values.

With such $Q/M$ values we see that each element of the original spherical
shell and the large mass $M$ is in balance with the electric and
gravitational forces neutralising one another.  But that is just the
condition required for the metric to be one of what Synge\cite{27}
calls conformastatic spaces, whose metrics take the delightfully
simple form
$$ds^2 = V^{-2} dt^2 - V^2 (dx^2 + dy^2 + dz^2) \ , \eqno (3.1)$$
where $V = V (x,y,z)$.

Take any function $V$ that $\rightarrow 1$ at infinity and satisfies
$V^{-3} \nabla^2 V \leq 0$.  
Then defining a proper energy density $\rho$ by
$$V^{-3} \nabla ^2 V = - {\textstyle {1 \over 2}} \kappa \rho \ ,
\eqno (3.2)$$
where $\kappa = 8 \pi G/c^4$ and
$$\nabla ^2 = {\partial^2 \over \partial x^2} + {\partial ^2 \over
\partial y^2} + {\partial^2 \over \partial z^2} \ ,$$ we find that the
metric solves the Einstein-Maxwell equations with electric field ${\bf
E} = (G)^{1/2} \mbox{\boldmath $\nabla$} V^{-1}$.  Here $\mbox{\boldmath
$\nabla$} = \left( {\partial \over \partial x}, {\partial \over  \partial y},
{\partial \over \partial z} \right)$.  Since these solutions are more
commonly described for arbitrary point masses with the appropriate
charges we give a brief outline of them in Appendix C.  Now $\nabla^2$
is an operator in $(x, y, z)$ space while $\rho$ is a proper energy
density of charged dust.  To find the corresponding coordinate density
of dust we set
$$\rho \, d  {(\rm Volume)} \ = \rho V^3 dxdydz = \rho_c dxdydz \ ,$$
so $\rho_c (x,y,z) = \rho V^3$,
hence $$\nabla ^2 V = - {\textstyle{1 \over 2}} \kappa \rho_c \ . \eqno (3.3)$$

For our problem we need the conformastatic metric of a spherical shell
with a distant charged mass, see Fig. 1.  Taking relativistic units $G
= c = 1$,
$$V = \left\{ {{M \over R} + {m \over r} + 1 \qquad r \geq b \atop {M
\over R} + {m \over b} + 1 \qquad r \leq b} \right \} \ , \eqno (3.4)$$
where $R$ and $r$ are the $x,y,z$ coordinate distances from $M$ and
$m$.  Far away from both shells, $g_{00} = V^{-2} = 1 - 2 M/R - 2m/r$.

Any test mass with charge to mass ratio $\sqrt{G}$ will be in static
equilibrium with balancing gravitational and electric forces if placed
at any point in such a metric.  However we shall be concerned with the
motion of an {\it uncharged} test mass.  Placed well away from the
spherical shell $m$ and at the same distance $Z$ from $M$, Figure 1, a
test mass will accelerate\footnote{Of course, an uncharged test mass
follows the geodesics and the invariant acceleration vanishes.  Here
we mean acceleration relative to the static frame invariantly defined
by the Killing field.} towards $M$ with initial acceleration $|{\bf g}|
= {M \over Z^2}$ but will the acceleration of a test mass placed {\it
inside} the spherical shell be greater or less than $|{\bf g}| $?

Relative to freely falling axes the shell is accelerated upwards by
the electric field so we expect a test mass inside to have an
acceleration in sympathy.  Thus relative to static axes we expect the
acceleration of the test mass within the sphere to be smaller than $|{\bf
g}| $.  However in strong field general relativity there are other
effects that work in the same direction.  What should we compare,
coordinate accelerations, $g_c = \left | d^2 {\bf r}/ d t^2 \right |$,
proper accelerations, $g_p = \left | d/d \tau \left( V d{\bf r} / d \tau
\right) \right|$, or proper length accelerations using universal
coordinate $t$-time, $g_u = \left | d/dt \left(V d{\bf r} / dt
\right) \right| $?  Here ${\bf r}$ stands for $(x,y,z)$.  One could
argue that because time runs slowly at greater gravitational
potentials we expect a slower apparent acceleration in $t$-time
irrespective of any sympathetic acceleration effect.  The equations of
geodesic motion in proper time $\tau$ are
$$d/d \tau (V^2 d{\bf r}/d \tau) = 
(2V^2 \varepsilon^2 - 1 ) \mbox {\boldmath $\nabla$} {\rm ln} V  \eqno (3.5)$$
and the conserved specific energy is $\varepsilon = V^{-2} {\dot t}$,
where a dot means $d/d \tau$.  Since $d \tau$ is $ds$ along the
particle's motion,
$$V^{-2} \, {\dot t}^2 - V^2 \, {\dot {\bf r}}^2 = 1 \ . \eqno (3.6)$$

From the above we find expressions for ${\bf g}_c \, , {\bf g}_p$ and
${\bf g}_u$.  For a body released from rest $(\varepsilon = V^{-1})$
the initial accelerations are
$${\bf g}_c = V^{-5} \mbox {\boldmath $\nabla$} V\ , \eqno (3.7)$$
$${\bf g}_p = V^{-2} \mbox {\boldmath $\nabla$} V\ , \eqno (3.8)$$
$${\bf g}_u = V^{-4} \mbox {\boldmath $\nabla$} V\ . \eqno (3.9)$$ Each
of these measures of acceleration is affected by the value of $V$ as
well as its gradient.  So, if the potential, $m/b$, due to the shell
is large, the acceleration of the body inside the shell will be much
reduced compared with that of another body far from $m$ but at the
same distance from $M$.  Whereas in (3.7) and (3.9) much of this
difference can be attributed to slowed time that does not account for
the result (3.8) where such effects are eliminated by using time as
measured on the particle.  In these conformastatic models there {\it
is} an invariant definition of the gravity field.  All are agreed that
the systems are in balance and that any test particle of charge to
mass ratio $\sqrt{G}$ can be placed anywhere in such a metric and
will be in equilibrium.  Thus the electric field in the static frame
(which is invariantly defined) must be equal and opposite to the
gravitational field ${\bf g}$.  Since the electric field is $\mbox
{\boldmath $\nabla$} (V^{-1})$, the gravitational field is $-\mbox
{\boldmath $\nabla$} (V^{-1}) = V^{-2} \mbox {\boldmath $\nabla$} V$.
So this invariant definition of the gravity field leads us back to
${\bf g}_p$.  Comparing those accelerations inside the shell to those
far from it we have
$$(g_p)_{{\rm in}}/(g_p)_{{\rm out}} = E_{{\rm in}} /E_{{\rm out}} = (V_{{\rm in}}/V_{{\rm
out}})^{-2} = \left( {1 + {M \over Z} \over 1 + {M \over Z} + {m \over
b}} \right)^2 \ , \eqno (3.10)$$ so if $m/b$ is larger than $1 + M/Z$
the reduction can be very large.  It is important to notice here that $b$
can be as small as we like; $b=0$ corresponds to an extreme Reinsner
Nordstr\"om black hole.  Indeed as shown by Hartle \& Hawking
1972\cite{28} our coordinate radius $r$ is related to Schwarzschild's
coordinate $r_s$ by $r = r_s - m$.

We have argued above that this reduction of the test mass's
acceleration is due to its sympathetic dragging produced by the
acceleration of the surrounding shell (relative to freely falling
axes).  However, let us now take the view of the static observer at
infinity; the shell is not accelerating at all.  It then appears
strange that this zero acceleration can have any effect.  Indeed
Einstein attributed such effects that did not involve his `vector
potential' ${\bf A}$ not to any sympathetic dragging but to mass
induction -- an increase of inertia due to the presence of the
surrounding sphere.  The calculation above shows that the reduced
gravity field, as seen in the {\em static} frame, is parallelled by
the reduction of the electric field.  We do not attribute the latter
to any change of charge, so the former should not be attributed to
Einstein's mass induction but is better viewed as a
``diagravitational'' effect of the potential, similar to a dielectric
effect, but reducing {\em both} the electric and the gravitational
fields inside the medium.  Evidently the sympathetic dragging
interpretation in the {\em freely falling} frame is equivalent to this
diagravitational effect in the static frame.

\section{Static Charge in a Conformastatic $(1 + \beta \cos \theta)$
Weighted Shell}

The method of the last section can be used to give a finite region of
almost uniform gravity field within a single spherical shell of total mass $M$ with all charges equal
to $\sqrt{G}$ times all masses.  Take
$$V = \left\{ {1 + {M \over r} - A b^3 r^{-2} \cos \theta \qquad r
\geq b \atop \ \ 1 + {M \over b} - A r \cos \theta \qquad \quad \ r
\leq b} \right\} \ \ , \eqno (4.1)$$ where $A$ is a constant.  Inside,
${\mbox{\boldmath $\nabla$}} V$ is constant but the gravity field $g$
varies at order $A^2$ as it is $V^{-2} {\mbox {\boldmath $\nabla$}}
V$.  The coordinate surface density of mass on the shell, $\sigma_c$,
is given by integrating (3.3) across the shell
$$4 \pi \sigma_c = 4 \pi V^{-2} \sigma = M b^{-2} - 3 A \cos \theta \
, \eqno (4.2)$$
where we have written $\sigma$ for the proper surface density per unit
proper area; the metric is 
$$ds^2 = V^{-2} dt^2 - V^2 \left[dr^2 + r^2 (d {\hat {\bf
r}})^2\right] = V^{-2} dt^2 - V^2 \left( dx^2 +dy^2 +dz^2 \right) \
. \eqno (4.3)$$ 
For $r \leq b$ the transformation $t = K{\overline
t}$, $K^{-1} = 1 + M/b$, $x = K {\overline x} (1 + {\overline A} \,
{\overline z})$, \break $y = K {\overline y} (1 + {\overline A}\, {\overline
z}), z = K \left[ {\overline z} + {1 \over 2} {\overline
A} \left( {\overline z}^2 - {\overline x}^2 - {\overline y}^2 \right)
\right], {\overline A} = K^2A$ yields for small $A$ $(A^2 b^2$
neglected)
$$ds^2 = \left( 1 + {\overline A} \, {\overline z} \right)^2 d
{\overline t}^2 - \left( d {\overline x}^2 + d{\overline y}^2 +
d{\overline z}^2 \right) \ , \eqno (4.4)$$ 
which is flat space in
accelerated axes (cf\cite{26}) and ${\overline A}$ is identified with
our former \break $a^{-1}$.  The physical electric field in our
conformastatic metric is ${\bf E} = \mbox {\boldmath $\nabla$} V^{-1}$
\break everywhere and this $\mbox {\boldmath $\nabla$}$ is the coordinate
$\mbox {\boldmath $\nabla$}$.  Writing it in terms of the physicists
gradient per unit distance $ \mbox {\boldmath $\nabla$}_p = V^{-1}
\mbox {\boldmath $\nabla$}$ so ${\bf E} = - \mbox {\boldmath
$\nabla$}_p \, {\rm ln} V$ and ln$V$ is the physicists potential (in
volts say).  Now this is invariant for purely spatial coordinate
changes like that above.  Hence the potential due to adding a small charge
$q$ at the origin (with the mass appropriate for the conformastat) is
$q/(rV)$.  Rewriting $r$ and $V$, $r = K {\overline r} \left(1 + {1
\over 2} {\overline A} \, {\overline z} \right)$ and $V = K^{-1}
\left( 1 - {\overline A} \, {\overline z} \right)$ we have the
potential to order ${\overline A} = 1/a$,
$$\Phi = {q \over {\overline r}} \left( 1 + {1 \over 2}\, {{\overline z} \over a} \right) \ , \eqno (4.5)$$
which agrees with Born's field in accelerated axes to this order.

The field ${\bf E}$ is still given in conformastatic coordinates by $${\bf E} = \mbox {\boldmath $\nabla$} \left( \delta {1 \over V} \right) = - \mbox {\boldmath $\nabla$} \left( q r^{-1} V^{-2} \right) \ . \eqno (4.6)$$
Outside the field is still static and given by (4.6) but of course the
external form of $V$ must be used there.

\section{Radiation and the Equivalence Principle}

In the last section we had a static charge in the gravitational field
of a shell.  It produces the Born field locally but no radiation. In
Section 2 we had an accelerated charged sphere in flat space which
produced a `Born' field with radiation at large distances.

Here we analyse the electromagnetic field of a particle which after
moving uniformly at high speed has decelerated to rest and will
continue its constant acceleration until it disappears whence it
came.  We draw the field lines and show that it radiates in a region
inaccessible to the coaccelerating observer which does not exist for
the static charge in the shell.  Then we consider the corresponding
problem for accelerated masses producing gravitational radiation.

Figure 2a is a space-time diagram with time plotted to the left and
$z$ upwards.  The reason for $z$ up will soon appear.  $t$ is plotted
leftwards so that creatures of habit can recover their familiar
space-time diagram by rotating the page by 90$^\circ$.  The charged
particle is confined to the line $x = y = 0$ and at $P$ it is moving
downwards at a little less than the light velocity.  However it is
uniformly accelerated upwards, for $t \geq t_0$, so that the
acceleration felt in the momentary rest frames is constant thereafter.
This acceleration slows it to momentary rest at $O$ before it continues
to accelerate upwards to $O'$ and beyond.  The space-time diagram is
drawn using coordinates and time appropriate for its rest moment at $O$.
Prior to arrival at $P$ the particle was moving downwards at constant
velocity so the acceleration only started at $P$.  It is clear from
Fig. 2a that a coaccelerated observer who moves with the particle can
never see anything in the lower left of the diagram below the shaded
line, ${\overline z} = z + a = t$.  Light from such points can never
reach any point of the hyperbola $POO'$.  Thus the region ${\overline
z} < t$ is beyond his horizon which is at $\overline z = t$.  Likewise
nothing done by such an observer can influence the region below
$\overline z = -t$ unless it was already done before the particle
started to accelerate (and before the particle crossed out of that
region at $S$).

Figure 2b shows a perpendicular slice of Minkowski space containing
the $x,z$ plane through $O$.  The lines of force of the Born electric
field are a set of circles through $O$ whose centres lie on $z = -a$,
(i.e., $\overline z = 0$).  $P'$ is the same point of space as $P$ but
at the later time (as reckoned in this Minkowski frame).  Beyond the
circle of radius $t_0$ around $P'$ there has not been time for the
field to feel the acceleration that started at $t_0$ so the field is
the old field that there would have been had the particle continued
its prior uniform motion to $Q$.  Now the field of a fast moving
particle is a radial electric field flattened by Lorentz contraction
along its direction of motion (See Fig. 2d) and accompanied by circles
of magnetic field around the direction of motion; so this is the field
centred on $Q$ beyond $t_0$ from $P'$.  The field lines of the Born
field in Figure 2b droop because in the accelerated frame that travels
with the particle there is an apparent gravity field downwards and
gravity acts on the field's energy and stresses.  There is no magnetic
field in the Born region of this diagram because we have taken the
time symmetric moment, $t = 0$, when the particle is momentarily at
rest.  Then the Born fields in both the Minkowski and the accelerated
frames coincide.  There is no transient between the Born field and the
fast particle's field, merely the circular front (see Appendix B).  It
is of interest to see how the field in Figure 2b evolves at later
times.  We draw it in two instructively different ways when the particle
has moved on to $O'$.  It seems natural to take moving axes and draw
the field in the frame in which the particle at $O'$ is at rest.  We
then get Figure 2c in which the Born part of the field extends
farther, the front is now closer to $X$ and the original fast
particle's field beyond it is more Lorentz contracted since in these
axes the particle at $P$ was moving faster.  It is instructive to see
what cut in space-time corresponds to this diagram.  It is $O'X$ in
Figure 2a.  Thus the set of such pictures in which the Born field is
static are a set of cuts through $X$.  These diagrams only cover the
upper and lower of the four sectors through $X$, the others are
unrepresented so far.

On the cut $PX$ the Born region has shrunk to zero and $P$ is reduced
to rest so the field is just that of a static charge.  We have not
drawn that figure.  However we have drawn in Figure 2d the field lines
of the rapidly moving unaccelerated charge corresponding to the slice
$YP$ through Figure 2a at time $t_0$.  This shows the Lorentz
contracted Coulomb field and the associated magnetic field in circles
about the motion.  The corresponding slice at time $t=0$ is of course
Figure 2b.  In Figure 2e we draw the vertical slice through $O'$.  We
now reach into the radiation zone invisible to coaccelerating
observers and indeed $Q$ has now moved into that zone at $Q''$.  The
electric field lines are again circles through $O'$ centred on
$\overline z = 0$ but now they are accompanied by magnetic fields in
circles and start to converge again before joining the old field of
the fast moving particle outside the circular front.  Some may feel
surprised that the fields change discontinuously on this front
although the potentials are of course continuous.  One might have
expected that discontinuous changes in field would always be
accompanied by surface densities of charge or current but in fact such
changes can happen on null surfaces as is the case here.  An even more
extreme configuration is illustrated in Gull {\it et al}.\cite{29a} among
their pictures of various electromagnetic fields.  To see the
radiation in the Born field one sets $t = u+r$ which brings one clearly
into the ``radiative zone'' of Figure 2a and takes the asymptotic
form at large $r$ with $u$ fixed.  This gives
$${\bf E} = qa^2 {{\hat {\bf r}} \times ({\hat {\bf r}} \times {\hat
{\bf z}}) \over r   (u^2 + a^2 \sin^2 \theta)^{3/2}}\ , \eqno (5.1)$$
$${\bf B} = qa^2 {{\hat {\bf z}} \times {\hat {\bf r}} \over r  (u^2 + a^2 \sin ^2
\theta)^{3/2}}\ . \eqno (5.2)$$
Thus ${\bf E}$ and ${\bf B}$ are perpendicular and perpendicular to
${\hat {\bf r}}$, and fall only as $r^{-1}$ which demonstrates the
radiative field in this region.

The only explicit solutions of Einstein's equations available today
that represent radiative gravitational fields of finite sources have
much in common with the Born solution in electrodynamics.  These
``boost-rotation symmetric'' space-times have two symmetries exhibited
by the axial Killing vector, and the boost Killing vector which, in
appropriate coordinates, has the flat-space form, $\zeta^{\alpha} =
(z, 0,0,t)$.  The Born solution has the same symmetries.  The
boost-rotation symmetric space-times represent the fields of
``uniformly accelerated particles'' which may be black holes as is the
case for the C-metric.  The ``sources'' of the accelerations are
strings along the z-axis.  Analogously to the Born solution,
gravitational waves are found in the region $|t| >|z|$ where the boost
Killing vector is space-like.  In the region $|z| > |t|$, the metrics
can be transformed into static forms -- the sources are at rest in
uniformly accelerated reference frames, again in full analogy with the
Born case.  The radiative character of the solutions is best seen on
asymptotic forms of the Riemann tensor at null infinity.  The Riemann
tensor falls-off as $r^{-1}$, in the analogy with the ``radiative
fall-off'' of electric and magnetic fields at $r \rightarrow \infty$
with $u$ fixed, given in (5.1), (5.2).  In addition, the analogous
features of fields due to some specific exact boost-rotation symmetric
space-times and uniformly accelerated charges appear even in the
radiation patterns and total rates of radiation, as first demonstrated
many years ago\cite{29}.

The structure of general boost-rotation symmetric space-times was
analysed much later by Bi\v{c}\'{a}k \& Schmidt (1989)\cite{30}.  In
the review ``Radiative space-times: exact approaches''\cite{31}, the
main properties and applications of the boost-symmetric space-times
are summarised.

\section*{Appendix A Born's Solution $(c = 1)$}

Born's solution is written in terms of $R^2 = x^2+y^2$ and $\overline
z = z + a$.  We put
$$\zeta = \left[ \left( a^2 + t^2 - R^2 - \overline z^2 \right)^2 +
4a^2 R^2 \right]^{1/2} \ . \eqno ({\rm A}1)$$
The Born potentials are 
$$\Phi^B = q [\overline z (R^2 + \overline z^2 + a^2 - t^2) \zeta^{-1}
- t ]/ (\overline z ^2 -t^2) \eqno ({\rm A}2)$$
$${\bf A}^B = {\hat {\bf z}} A^B = {\hat {\bf z}} q [t (R^2 +
\overline z^2 +a^2 -t^2 ) \zeta^{-1} - \overline z]/(\overline z^2 - t^2) \
. \eqno ({\rm A}3)$$
From these follow the fields 
$${\bf E}^B = 4a^2 q \zeta^{-3} [- (a^2 +t^2 +R^2-{\overline z}^2) {\hat {\bf z}}
+ 2R{\overline z} {\hat {\bf R}}] \ , \eqno ({\rm A}4)$$
$${\bf B}^B = 8a^2 q\zeta^{-3} Rt \, {\hat {\bf z}} \times {\hat {\bf R}} \ ,  \eqno ({\rm
A}5)$$
where ${\hat {\bf R}}$ and ${\hat {\bf z}}$ are unit vectors along
$(x,y,0)$ and $(0,0,1)$ respectively.  

To evaluate the dipolar contribution to (2.17) we need
$$(\partial \zeta^{-1}/\partial a^2)_{\overline z} = - \zeta^{-3} (a^2
+ t^2 + R^2 - \overline z^2) \eqno ({\rm A}6)$$
$$(\partial \Phi^B/\partial a^2)_{\overline z} = -2q \overline z
\zeta^{-3} (t^2 +a^2 - R^2 - \overline z^2) \eqno ({\rm A}7)$$
$$\partial A^B /\partial a^2 = (ct/\overline z) \, \partial
\Phi^B/\partial a^2 \ . \eqno ({\rm A}8)$$

Since dipoles are generated by small shifts we have following
Bi\v{c}\'{a}k \& Muschall (1980)\cite{25}
$$\Phi = \Phi^B +{{\bf P} \cdot {\hat {\bf z}} \over Q} \left({
\partial \Phi^B \over \partial a} \right)_{\overline z} = \Phi ^B +
{\textstyle {1 \over 3}} b^2 \left( {\partial \Phi^B \over \partial
a^2} \right) _{\overline z} \eqno ({\rm A}9)$$
with an analogous formula for ${\bf A}$.  Thus the total external
potential $\Phi$ is given by multiplying (A7) by ${\textstyle {1 \over
3}} b^2$ and adding it to (A2).  ${\bf A}$ is constructed similarly.

\section*{Appendix B  Electromagnetic Transients}

It is of interest to see how the fields arise when the sphere is at
rest initially and is then uniformly accelerated from $t=0$ onwards.
To find such solutions, we put $D = {\textstyle {1 \over 2}} \alpha
t^2 \Theta (t)$ and evaluate our formulae as before.  We find
from (2.10) and (2.13) that for $r \leq b$, $\Phi$ takes the values
$$\left\{ \begin{array}{ll}
\! \! \! Qb^{-1} & \! \! \! \ \,  t \leq {b - r \over c} \\
\! \! \! Qb^{-1} \left\{ 1 + {\textstyle {1 \over 12}} \alpha cz r^{-3} \left( -3 {r
\over c} + t + {r-b \over c} \right) \left( t + {r-b \over c}
\right) ^2 \right\} & \! \! \! \ \, {b + r \over c} \geq t \geq {b - r \over c} \\
\! \! \! Qb^{-1} \left( 1 - {\textstyle {1 \over 3}} \alpha z c^{-2} \right) &
\! \! \! \  \, t \geq {b + r \over c}
\end{array} \! \! \! \right \} $$
and
$${\bf A} = \left\{ \begin{array}{ll}
 0 & \ \ \ t \leq {b - r \over c} \\
 - {\textstyle{1 \over 4}} Q b^{-1} {\hat {\bf z}} \alpha r^{-1} \left( t +
{r - b \over c} \right)^2 & \ \ \ {b +r \over c} \geq t \geq {b - r \over c}
\\ 
-Qb^{-1} {\hat {\bf z}} \alpha c^{-2} (ct -b) & \ \ \ t \geq {b +r \over c}
\end{array} \right\} $$
while for $r \geq b$, $\Phi$ takes the values
$$\left\{ \begin{array}{ll} Qr^{-1} & \ \ \ t \leq {r-b \over
c} \\ Qr^{-1} \left\{ 1 - {\textstyle {1 \over 12}} \alpha cz b^{-1}
r^{-2} \left( {3r \over c} + t - {r-b \over c} \right) \left( t - {r-b
\over c} \right) ^2 \right\} & \ \ \ {r+b \over c} \geq t \geq  {r -b
\over c} \\ Qr^{-1} \left\{ 1 - {\textstyle {1 \over 2}} \alpha 
c^{-2} z \left[ 1 - { (3c^2 t^2 +b^2) \over 3r^2} \right] \right\} & \ \ \ t \geq {r +b \over c} 
\end{array} \right \}$$
and
$${\bf A} = \left\{ \begin{array}{ll}
0 & \ \ \ t \leq {r - b \over c} \\
- {\textstyle{1 \over 4}} Q r^{-1} {\hat {\bf z}} \alpha b^{-1} \left( t -
{r - b \over c} \right)^2 & \ \ \ {r +b \over c} \geq t \geq {r - b \over c}
\\
-Qr^{-1} {\hat {\bf z}} \alpha c^{-2} (ct -r) & \ \ \ t \geq {r +b \over c}
\end{array} \right\} \ . $$

Evidently the internal potentials and fields settle down to their
steady state after a transient that lasts for a period of $2b/c$.
However these transients travel outwards with the velocity of light
and are seen later at greater distances as we would expect.
Transients for other starting times can be found by the Lorentz
transformation which reduces the starting velocity to rest.  We may
apply similar considerations to the case $b = 0$, a point charge
started from rest.  The field of a static or uniformly moving charge
changes to Born's field with a transient that lasts for zero time at
each point and travels outward with the velocity of light.

\section*{Appendix C  Conformastats}

These static space-times, so named by Synge\cite{27}, have conformally
flat spatial metrics. The Schwarzschild solution is a conformastat but
the best known examples are electrovacuum solutions due to
Majumdar\cite{32} and Papapetrou\cite{33} representing extreme
Reissner-Nordstr\"om black holes in arbitrary positions in static
equilibrium.  Less known (not explicitly included in\cite{34}) are the
conformastatic space-times with charged dust, in which the ratio of
the charge to mass densities $\rho_ e/\rho$, is everywhere $\pm
\sqrt{G}$.  (In Section 3, we took $\rho _e/\rho = +\sqrt{G}$.)
  As in Newtonian physics the material is in equilibrium because the
mutual gravitational attractions are balanced by the electrical
repulsions.  This class of exact solutions was studied, for
example, by Das\cite{35}, Bonnor and Wikramasuriya\cite{36}, and
Bonnor\cite{37}.  We did not, however, find a clear simple derivation
of the general solutions of this type.  Following Synge\cite{27},
consider the particular conformastat form of the metric, 
$$ds^2 = V^{-2} dt^2 - V^2 (dx^2 + dy^2 + dz^2) \ , \eqno ({\rm C}1)$$
where $V = V (x,y,z)$.  As a source, take a static electric field
described by potential $\phi (x,y,z)$ and a charged dust with static
proper matter and charge densities $\rho (x,y,z)$ and $\rho_e (x,y,z)$
(not initially assumed to be equal), in the coordinates of the metric
(C1).

Therefore, the energy-momentum tensor of the dust is $T^{\mu \nu}\,
({\rm dust}) \ = \rho
U^\mu U^\nu$, and its current density is $j^\mu = \rho_e U^\mu$, where
$U^\mu = \delta ^\mu_0 V$.  The electromagnetic field is described by
$F_{\mu \nu} = A_{\nu , \mu} - A_{\mu , \nu}$, with
$A_\mu = (\phi, 0,0,0)$, and the standard energy-momentum tensor
$$T_{\mu \nu}({\rm em}) \ = {1 \over 4 \pi} \left( - F_\mu ^\sigma F_{\nu
\sigma} + {\textstyle {1 \over 4}} g_{\mu \nu} \, F_{\rho \sigma} F
^{\rho \sigma} \right) \ . $$

The Ricci tensor components for the metric (C1) read
$$R_{ij} = -2 V^{-2} V,_i V,_j + \delta_{ij} V^{-1} (- \nabla^2 V +
V^{-1} \nabla V \nabla V ) \ , \eqno ({\rm C}2)$$
$$R_{00} = V^{-5} (V^{-1} \nabla V \nabla V - \nabla^2 V) \ , \eqno
({\rm C}3)$$
$$R_{0i} = 0 \ , \eqno ({\rm C}4)$$
the scalar curvature is
$$R = 2V^{-3} \nabla^2 V \ . \eqno ({\rm C}5)$$
Here the standard flat-space notation is used:
$$\nabla V \nabla V = V,_i V,_i \, , \, \nabla^2 V = \left
( {\partial^2 \over \partial x^2} + {\partial^2 \over \partial y^2} +
{\partial^2 \over \partial z^2} \right) V \ . $$
(Our Ricci tensor components are defined so that they have opposite
signs to those of Synge\cite{27} in his equations (181) and his $U$ is
our $V$.)  Now the non-diagonal space components of the Einstein
equations,
$$-2 V^{-2} V,_i V,_j = {\kappa \over 4 \pi} (-V^2 \phi,_i \phi,_j), \qquad \kappa = 8 \pi
G \ , \eqno  ({\rm C}6)$$
imply
$$\phi = - {1 \over \sqrt{G} V} \ , \eqno ({\rm C}7)$$ where we have
chosen the sign that corresponds to positive charges and we have set
to zero an additive constant in the electric potential.  [The solution
with $\phi = + {1 \over \sqrt{G}V}$ may be obtained by altering the
signs of $\phi, {\bf E} $ and $\rho_e$.]  Using $T_{ij} \, ({\rm
dust}) \ = 0$ for $i = j$ and $T_{ij} ({\rm em}) \ = {1 \over 4 \pi}
\left( -V^2 \phi,_i \phi,_j + {\textstyle{1 \over 2}} \delta_{ij} V^2
{\mbox {\boldmath $\nabla$}} \phi \cdot {\mbox {\boldmath $\nabla$}}
\phi \right) $ for $i=j$ and substituting for $\phi$ from (C7), we see
that the diagonal spatial components of the Einstein equations are
automatically satisfied.  Now the last Einstein equation to be solved,
$R_{00} - {\textstyle{1 \over 2}} g_{00} R = \kappa \left(T_{00} \ ({\rm
dust}) \ + T_{00} \ ({\rm em})\right)$ implies the non-linear Poisson-type
equation
$$\nabla^2 V = - {\textstyle{1 \over 2}} \kappa \rho V^3 \ , \eqno ({\rm C}8)$$
while the sole non-trivial Maxwell equation to be satisfied, is
$$V^{-2} {\mbox {\boldmath $\nabla$}}\ . \ (V^2 {\mbox {\boldmath
$\nabla$}} \phi) = -4 \pi \rho_e V \ , \eqno ({\rm C}9)$$
with $\phi$ given by (C7) and $V$ fulfilling (C8).  By comparison we
conclude that the charge density must necessarily be equal to the mass
density or, in non-relativistic units,
$$\rho_e = + \sqrt{G} \rho \ . $$

For {\it any} function $V (x,y,z)$ which satisfies $V^{-3} \nabla^2 V
\leq 0$, Eq. (C8) guarantees that the mass density is non-negative.
Thus, a solution describing space-time with an arbitrary mass density
$\rho = -2 \kappa^{-1} V^{-3} \nabla^2 V$, which is kept in
equilibrium by the charged density $\rho_e = \sqrt{G} \rho$, is
constructed.  The additional condition, $V \rightarrow 1$ at infinity,
makes the space-time asymptotically flat.  The physical (frame)
components of the electric field are given by
$${\bf E} = - {\mbox {\boldmath $\nabla$}} \phi = {1 \over \sqrt{G}}
{\mbox {\boldmath $\nabla$}} V^{-1} \ . \eqno ({\rm C}10)$$
These conformastats are useful because one can construct bodies of
arbitrary shape and mass density\cite{36}, or study the model problems
as here in Sections 3 and 4.

\newpage

\begin{figure}[th]
\centerline{\psfig{figure=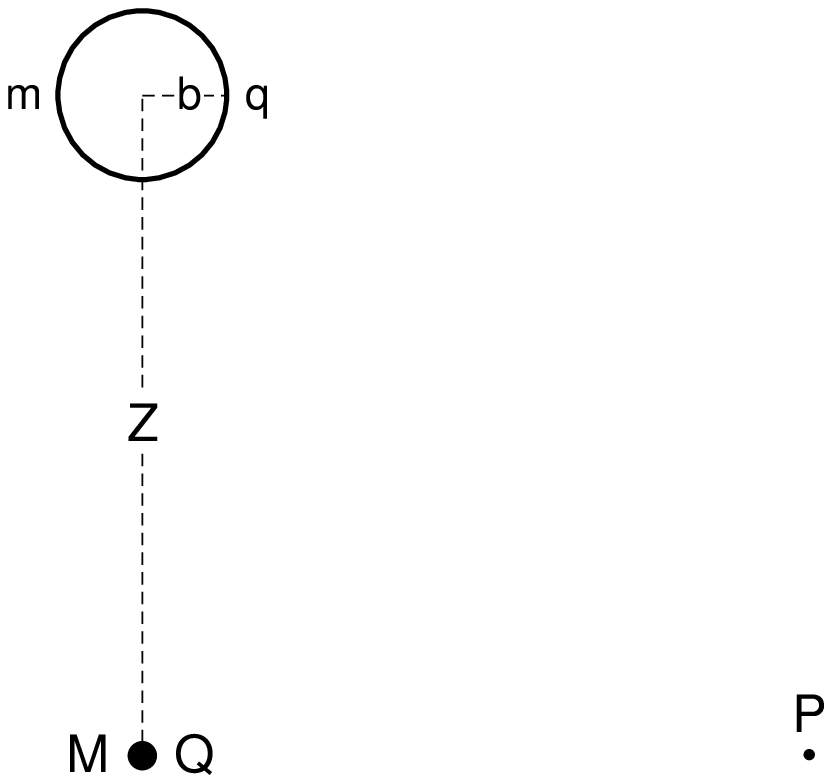,height=3in}}
\caption{A spherical shell of radius $b$, mass $m$ and charge $q$ has
a significant gravitational potential inside.  It lies at coordinate
distance $Z$ above a great mass $M$ with charge $Q$.  They are of
equal charge-to-mass-ratio, $\sqrt{G}$, and remain in equilibrium with
the gravitational attractions balancing electrical repulsions.  The
acceleration of an uncharged test mass within the shell is compared to
that at $P$ also distant $Z$ from $M$ but far from the shell $m$.}
\end{figure}
\newpage

\renewcommand{\thefigure}{2a}
\begin{figure}[th]
\centerline{\psfig{figure=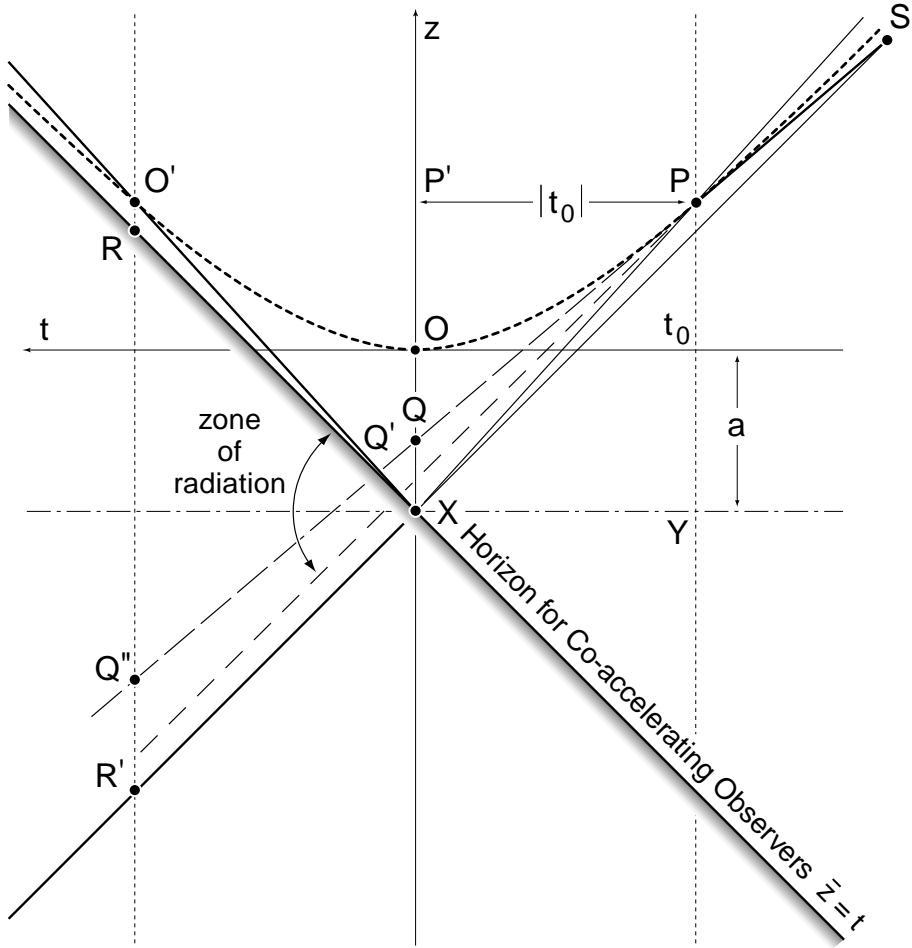,height=5in}}
\caption{In this space-time diagram the $z$-axis is upwards and time
increases to the left.  A charged particle moves uniformly downwards
till it reaches $P$ when it is uniformly accelerated upwards coming to
momentary rest at $O$ before accelerating up to $O'$ and beyond.  Had
the particle not been accelerated it would have followed $SPQQ'Q''$.
The shaded line $\overline z = t$ is the event horizon for an observer
accelerating with the charge.  $X$ is a distance $c^2/\alpha = a$
below $O$; $\alpha$ is the acceleration.}
\end{figure}

\newpage

\renewcommand{\thefigure}{2b}
\begin{figure}[th]
\centerline{\psfig{figure=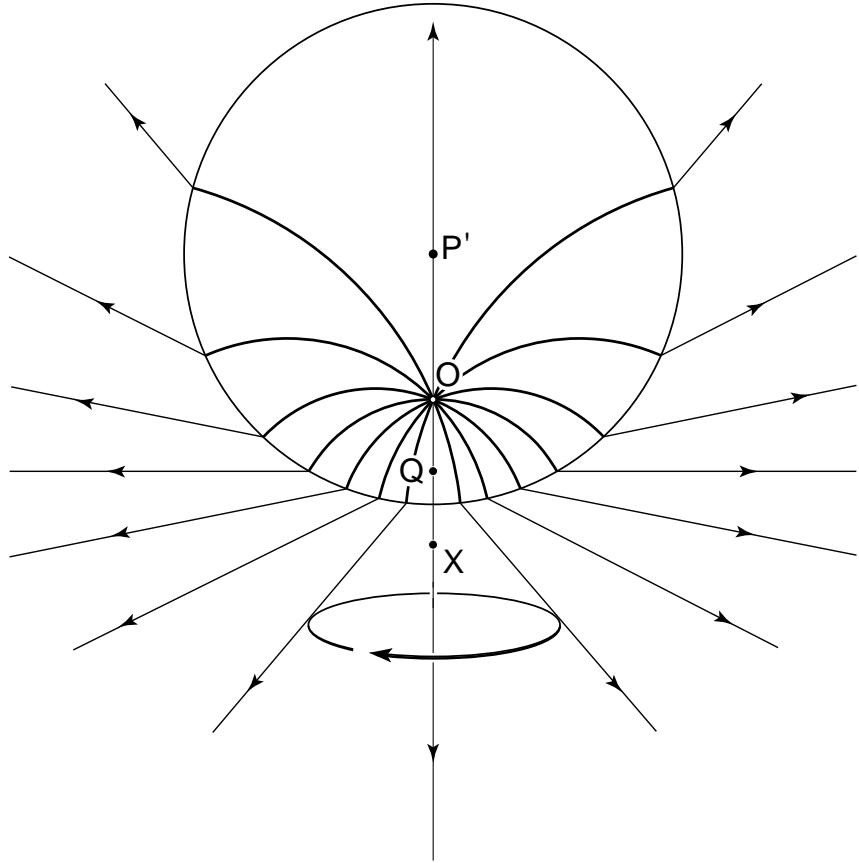,height=4.5in}}
\caption{Shows the lines of electric force in the $xz$ plane that
corresponds to the line $OX$ through Figure 2a.  The particle is
momentarily at rest at $O$ but accelerating upward.  The field lines
droop under the effective gravity induced by this acceleration but
beyond $ct_0$ from $P'$ there has not been time for the field to be
modified from what it would have been had the particle continued on
its unaccelerated path to $Q$.  The field would then have pointed at
the particle so outside a sphere of radius $ct_0$ it still points there.
There is no magnetic field within that sphere but outside it the field
has a magnetic component as though the particle was at $Q$ and
still moving downward at a rate that flattens the Coulomb field by a
factor 2.  The magnetic field is only present in this outer region and
lies in rings about the particle's original motion.}
\end{figure}

\newpage

\renewcommand{\thefigure}{2c}
\begin{figure}[th]
\centerline{\psfig{figure=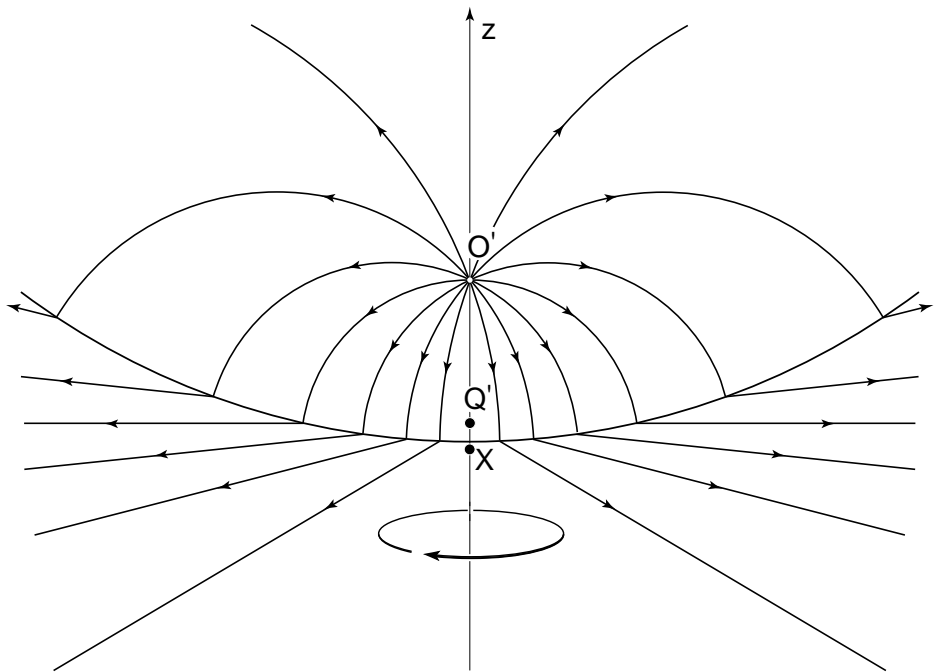,height=3.5in}}
\caption{This corresponds to the slice $O'X$ through Figure 2a.  The
electric field lines are again shown after a Lorentz transformation
has reduced the particle at $O'$ to rest.  Had the particle continued
uniformly in a straight line from $P$ it would have been at $Q'$, the
far field still points there since the message that the particle has
been accelerated has not yet come there.  In this frame the particle
was originally moving even faster so its Coulomb field is flattened by
a factor 4.}
\end{figure}

\newpage

\renewcommand{\thefigure}{2d}
\begin{figure}[th]
\centerline{\psfig{figure=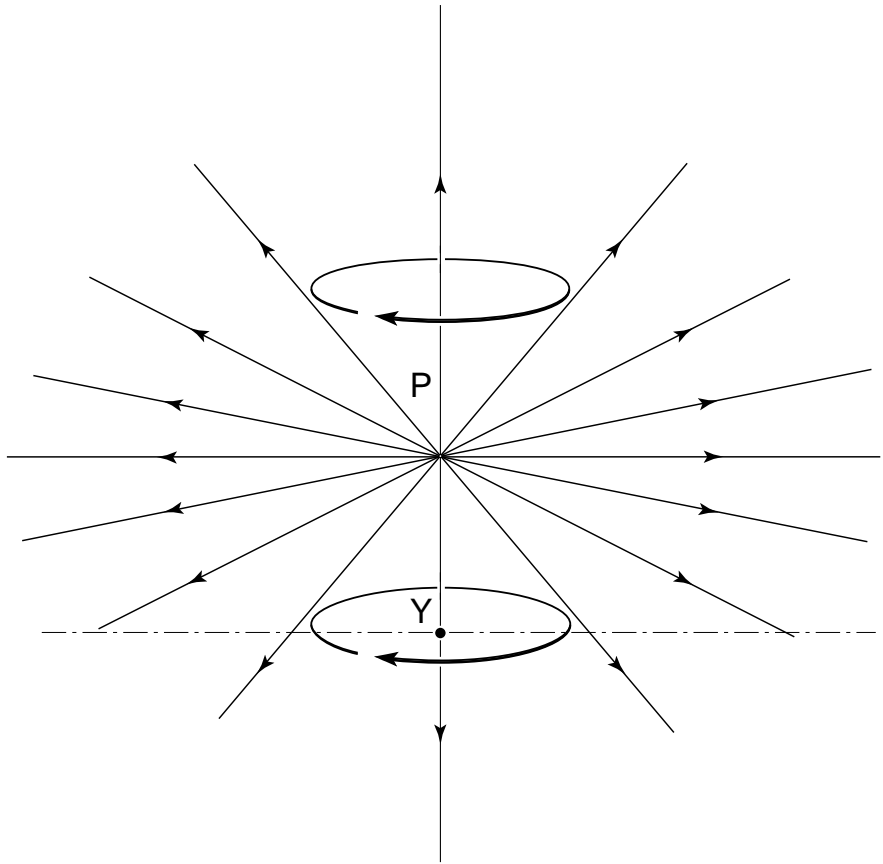,height=4.5in}}
\caption{This corresponds to the vertical slice $PY$ through Figure
2a at time $t_0$.  The electric field lines are those of a rapidly
moving charged particle with the Coulomb field flattened by a factor 2
and the magnetic field of a downward moving charge.}
\end{figure}

\newpage

\renewcommand{\thefigure}{2e}
\begin{figure}[th]
\centerline{\psfig{figure=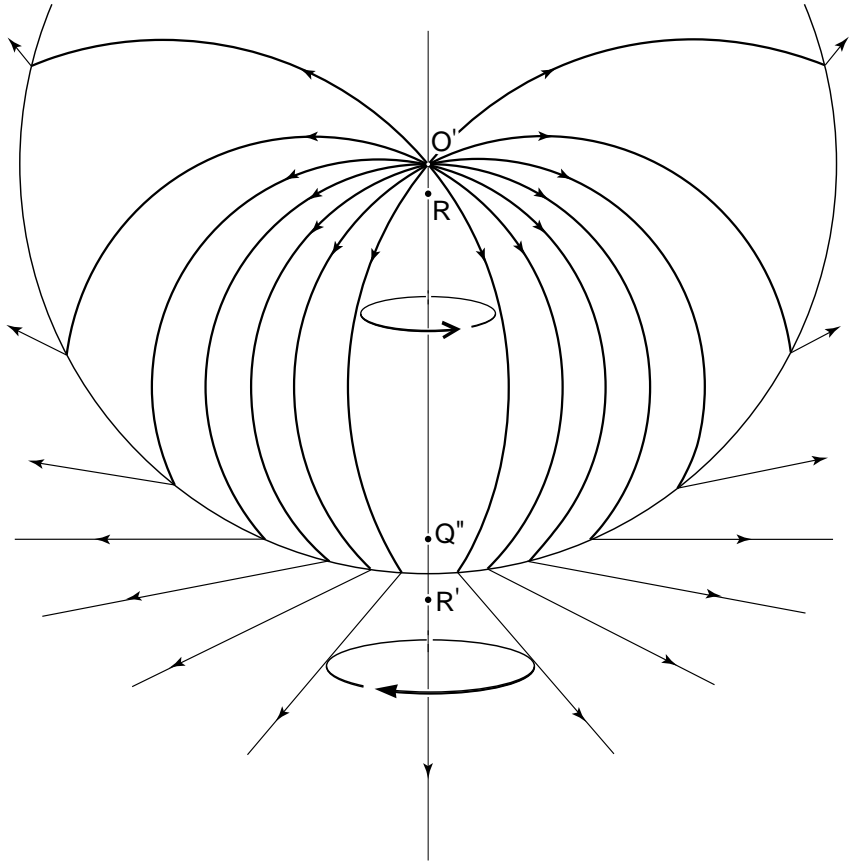,height=4.5in}}
\caption{This corresponds to the vertical slice $O' Q''$ through
Figure 2a.  The drooping electric field lines even start reconverging
in the region beyond the particle's event horizon which is the zone
where radiation is found.  In the near zone the magnetic field has
felt the upward motion of the particle so lies in opposing loops to
the outer zone where the field still does not appreciate that the
particle has ceased its downward motion.}
\end{figure}

\end{document}